\documentclass[conference]{IEEEtran}
\IEEEoverridecommandlockouts
\usepackage{cite}
\usepackage{amsmath,amssymb,amsfonts}
\usepackage{algorithmic}
\usepackage{graphicx}
\usepackage{textcomp}
\usepackage{xcolor}
\def\BibTeX{{\rm B\kern-.05em{\sc i\kern-.025em b}\kern-.08em
    T\kern-.1667em\lower.7ex\hbox{E}\kern-.125emX}}
\usepackage{multirow}
\begin{document}

\title{CAHS-Attack: CLIP-Aware Heuristic Search Attack Method for Stable Diffusion
}


\author{
Shuhan Xia$^{1}$\textsuperscript{*}\thanks{\textsuperscript{*}Equal contribution.} \quad
Jing Dai$^{2}$\textsuperscript{*} \quad
Hui Ouyang$^{3}$\textsuperscript{*} \quad
Yadong Shang$^{2}$ \quad
Dongxiao Zhao$^{3}$ \quad
Peipei Li$^{1}$\textsuperscript{\dag}\thanks{\textsuperscript{\dag}Corresponding author. Email: lipeipei@bupt.edu.cn} \\
$^{1}$ Beijing University of Posts and Telecommunications, Beijing, China \\
$^{2}$ China Mobile Ltd, Department of Cyber and Information Security Management, Beijing, China \\
$^{3}$ Aspire Information Technology (Beijing) Company Limited, Beijing, China \\
}

\maketitle

\begin{abstract}
Diffusion models exhibit notable fragility when faced with adversarial prompts, and strengthening attack capabilities is crucial for uncovering such vulnerabilities and building more robust generative systems. Existing works often rely on white-box access to model gradients or hand-crafted prompt engineering, which is infeasible in real-world deployments due to restricted access or poor attack effect. In this paper, we propose CAHS-Attack , a CLIP-Aware Heuristic Search attack method. CAHS-Attack integrates Monte Carlo Tree Search (MCTS) to perform fine-grained suffix optimization, leveraging a constrained genetic algorithm to preselect high-potential adversarial prompts as root nodes, and retaining the most semantically disruptive outcome at each simulation rollout for efficient local search. Extensive experiments demonstrate that our method achieves state-of-the-art attack performance across both short and long prompts of varying semantics. Furthermore, we find that the fragility of SD models can be attributed to the inherent vulnerability of their CLIP-based text encoders, suggesting a fundamental security risk in current text-to-image pipelines.


\end{abstract}


\section{Introduction}
\label{sec:intro}

In recent years, advances in text-to-image generation have led to the emergence of powerful models such as Stable Diffusion (SD)\cite{rombach2022high, stablility2023stable}, FLUX\cite{flux2024flux}, and MMaDA\cite{yang2025mmada}, enabling users to create high-quality images from natural language prompts.

\begin{figure}[t]
    \centering
\includegraphics[width=1.0\linewidth]{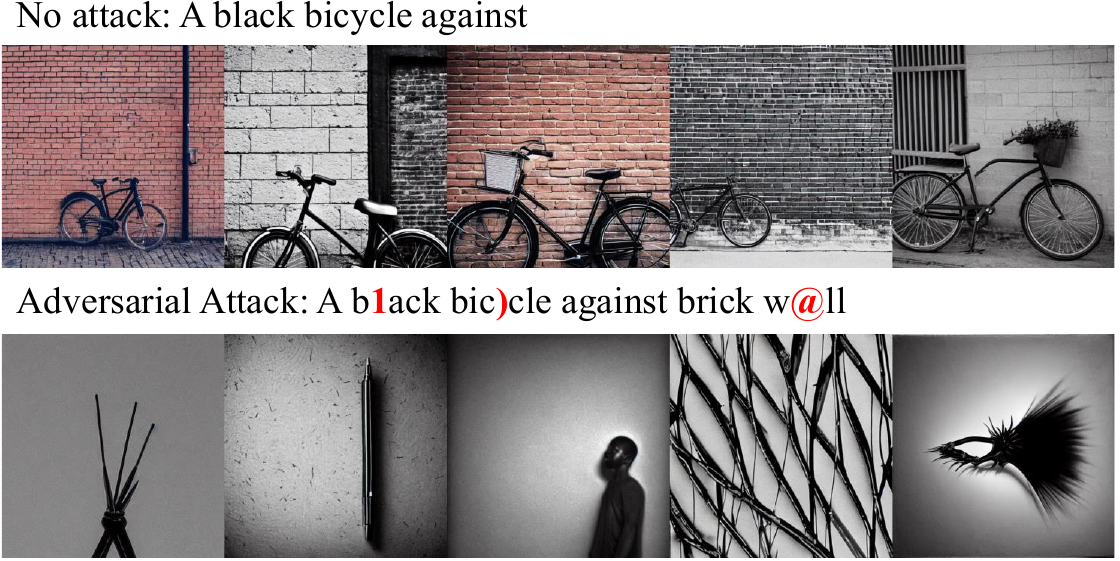}
    \caption{Illustrating the Fragility of Stable Diffusion under Adversarial Prompt Perturbations. The attacked characters are highlighted in \textcolor{red}{red}. }
    \label{fig:fig1}
\end{figure}

However, the incorporation of natural language introduces a new layer of vulnerability. Text prompts, while expressive, are inherently ambiguous and sensitive to subtle variations. As shown in Fig.~\ref{fig:fig1}, this can lead to severe inconsistencies in the output — for instance, a generated image may omit key subjects\cite{zhuang2023pilot}, misrepresent attributes\cite{du2023stable}, or produce toxic images\cite{yang2024mma, huang2025perception}. These issues not only reduce the reliability of generative models but also pose potential risks in real-world deployments, such as prompt-based manipulation or content injection.
Most existing adversarial attacks against Stable Diffusion (SD) treat it as a white-box system, relying on gradient information to optimize adversarial prompts\cite{ma2024jailbreaking, yang2024mma, jia2024improved, huang2024personalization}. Prior white-box attacks such as JPA\cite{ma2024jailbreaking} and MMA-Diffusion\cite{yang2024mma} utilize gradient-based optimization and token-level selection strategies to construct effective adversarial prompts in the discrete text space. However, in real-world scenarios, attackers often lack access to model gradients. Furthermore, many modern defense mechanisms\cite{liu2024latent, yang2024guardt2i} incorporate gradient obfuscation techniques to prevent attackers from obtaining usable gradients. Recently, several studies\cite{yang2024sneakyprompt, le2022perturbations, liu2025token, gao2024hts} have attempted to attack SD via naive prompt engineering under black-box settings. Yet, such approaches are often inefficient and yield limited attack performance. 

To address these limitations, we introduce a black-box attack method, CAHS-Attack, which implements  heuristic search within CLIP embedding space. CAHS-Attack integrates Monte Carlo Tree Search (MCTS) to construct adversarial prompts by appending heuristic-guided suffix characters. Unlike conventional MCTS, our approach first employs a constrained genetic algorithm to globally perturb the base prompt, selecting high-potential adversarial candidates as root nodes for tree expansion. To further enhance search efficiency, we retain the most semantically disruptive outcome from each MCTS simulation, enabling the discovery of locally optimal adversarial suffixes. Extensive experiments demonstrate that CAHS-Attack achieves state-of-the-art attack performance on both short and long prompts across diverse semantic domains, and further reveal that the vulnerability of SD models largely stems from the intrinsic fragility of their CLIP-based text encoders.

To summarize, our contributions are threefold:

\begin{enumerate}

\item We introduce CAHS-Attack, a novel heuristic search–based adversarial attack method which reveals that the vulnerability of Stable Diffusion stems fundamentally from the intrinsic fragility of its CLIP-based text encoder.

\item CAHS-Attack integrates Monte Carlo Tree Search (MCTS) to perform suffix-level heuristic optimization, where the root nodes are selected via a constrained genetic algorithm, and the most semantically disruptive result is preserved at each simulation rollout to ensure optimal adversarial suffix selection.
\item We conduct extensive experiments across various prompt lengths and semantic categories. Our attack method achieves state-of-the-art attack performance on both ImageNet-Short and ImageNet-Long datasets.

\end{enumerate}

\section{Related Work}

\subsection{Diffusion Model}
Diffusion models have become a leading class of generative models, offering high-quality and diverse outputs by learning to reverse a Markovian noising process. The foundational formulation, known as Denoising Diffusion Probabilistic Models (DDPM)\cite{ho2020denoising}, progressively transforms Gaussian noise into data samples through a learned denoising network. This framework has since been extended and refined in both efficiency and scalability. Diffusion models have demonstrated strong performance across various domains, including image generation\cite{wang2024instancediffusion}, image inpainting\cite{corneanu2024latentpaint}, 3D generation\cite{voleti2024sv3d}, and video synthesis\cite{xie2025progressive}. Their flexibility in conditioning mechanisms and sampling strategies makes them a versatile backbone for many generative tasks.



\subsection{Adversarial prompt attack in text-to-image models}
A few studies have shown that text-to-image diffusion models are highly sensitive to prompt perturbations, revealing vulnerabilities in their reliance on natural language conditioning\cite{wen2023hard}. 
Recent studies\cite{xiong2025prompt, guo2024cold} explored how textual perturbations propagate through CLIP to affect downstream generation. However, the attack efficacy remains limited, and the method does not examine how small, semantically controlled edits to the prompt itself—without structural alterations—can influence diffusion outputs. In contrast, our work also perturbs the prompt space, offering a more effective and interpretable pathway to probe the fragility of text-conditioned generative models.

\subsection{Black-box optimization for adversarial attack}
Recent works have explored black-box adversarial attacks on diffusion-based generative models, particularly under realistic query-limited or API-only settings. For example, Inception \cite{zhao2025inception} targets toxic prompts by introducing a recursive decomposition strategy that incrementally subdivides unsafe expressions into minimally triggering segments. MDEA \cite{chen2025multidiffeditattack} combines large language models with genetic search to craft adversarial prompts that restructure sensitive lexical elements. Despite these advancements, such black-box strategies still struggle to achieve high attack efficacy in unrestricted settings.

\section{Proposed Method}
\begin{figure*}[t]
    \centering
\includegraphics[width=1.0\linewidth]{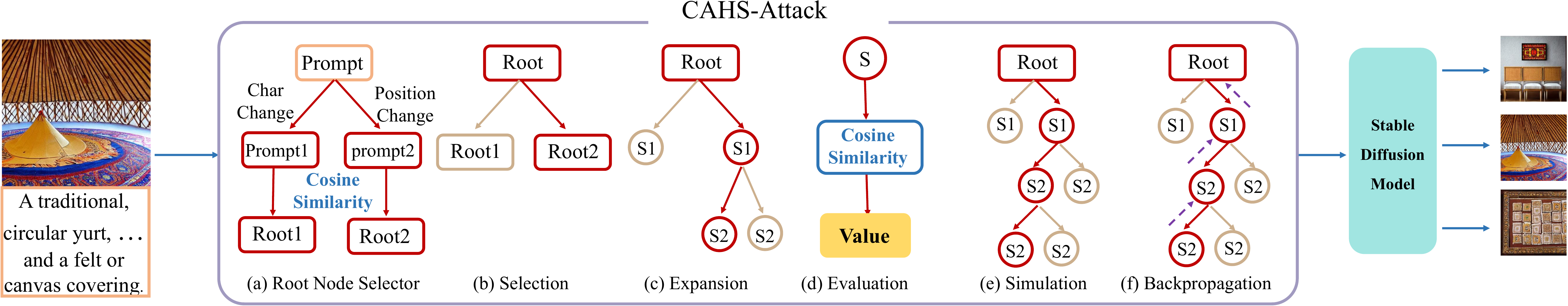}
    \caption{Overview of the six steps in CAHS-Attack. We begin with applying a constrained mutation-based strategy (left) to generate semantically plausible yet adversarially potent candidates as root nodes for MCTS. The subsequent MCTS process (middle) iteratively expands suffixes through selection, expansion, evaluation, simulation, and backpropagation, guided entirely by cosine similarity in the CLIP embedding space. The final adversarial prompt is then used to query the SD (right)}
    \label{fig:method}
\end{figure*}

In this section, we first provide an overview of Stable Diffusion, followed by a description of our attack model, which involves applying subtle perturbations to the text prompt in order to induce a significant semantic misalignment between the generated image and the clean prompt. Finally, we present the details of our proposed heuristic search attack method, CAHS-Attack. The pipeline of our attack method is illustrated in Fig.\ref{fig:method}.

\subsection{Problem Statement}

Stable Diffusion couples an encoder--decoder $(\mathcal{E}:\mathcal{X}\!\to\!\mathcal{Z},\ \mathcal{D}:\mathcal{Z}\!\to\!\mathcal{X})$ with latent-space diffusion. For text-to-image synthesis, a CLIP-based conditional encoder $\tau_{\theta}$ maps a prompt $c$ to embeddings that condition the denoiser $\epsilon_{\theta}$ via cross-attention. Starting from Gaussian noise $z_{T}$, the model iteratively applies
$\epsilon_{\theta}\bigl(z_{t},\, t,\, \tau_{\theta}(c)\bigr)$ to reduce noise and obtain a clean latent $z$, which is then decoded to the image $\mathbf{x}=\mathcal{D}(z)$. We observe that the text input in Stable Diffusion is entirely determined by the CLIP encoder. Therefore, perturbations to the text prompt directly affect certain feature dimensions within the CLIP embedding space.

\subsection{Attack Model}
We assume a threat model in which the attacker has access to the pretrained text encoder $\tau_\theta(\boldsymbol{c}) $ of CLIP, but not the internal parameters or gradients of the Stable Diffusion model. Let 
$x$ denotes the clean prompt and the $\boldsymbol{x}^{\prime}$ denotes perturbed prompt. Our attack aims to induce significant semantic deviation in the generated image by maximizing the difference in the CLIP text embeddings of $x$ and $x$. This objective can be formalized by minimizing the cosine similarity between the two embeddings, as Eq. \eqref{Equ:cos}.
 \begin{equation}
\begin{split}
\min_{\boldsymbol{x}^{\prime}}\cos(\tau_\theta(\boldsymbol{x}),\tau_\theta(\boldsymbol{x}^{\prime})),
\end{split}
\label{Equ:cos}
\end{equation}
where $\cos(\cdot,\cdot)$ denotes the cosine similarity. 

\subsection{CAHS-Attack}
\textbf{Root Node Selector.}
Traditional MCTS requires randomly selecting a root node to initiate the search. However, in the context of full-text prompt perturbation, the number of possible initial root candidates is combinatorially large, resulting in substantial computational overhead and reduced search efficiency. Although Genetic Algorithms (GA) is suitable for high-dimensional search, naive crossover destroys coherence and unconstrained mutation introduces noise, causing the adversarial prompts to deviate from the original text in terms of language and semantics.
To enhance the stealth of attack, we introduce a modified Genetic Algorithm with Fixed Perturbation Scope, designed for constrained and efficient perturbed prompt generation. GA eliminates the crossover operator and design a dual-mode mutation operator that restricts mutations to a bounded number of characters. 



Specifically, in each iteration, a fixed number $k$ of characters are selected for mutation. Let $x=\{x_1, x_2, ..., x_n\}$ denote the clean prompt, and $x' = \{x_1', x_2', ..., x_n'\}$ its perturbed counterpart. Mutation is applied only to positions $i \in \mathcal{V}$, where $|\mathcal{V}| = k$, by replacing $x_i$ with a sampled character $c_i$ from a predefined set $V$, as defined in Eq.~\eqref{Equ:mut}.
\begin{equation}
\begin{split}
\mathcal{M}(x) = \{x_i \mid i \in \mathcal{V}\}, \quad |\mathcal{V}| = k,
\end{split}
\label{Equ:mut}
\end{equation}
We evaluate the quality of each perturbation by computing the cosine similarity between the clean and perturbed prompt embeddings, and retain the top-$K$ perturbed prompt with the lowest similarity scores for the next generation.





To maintain a constant perturbation budget while ensuring sufficient exploration, we design a dual-mode mutation operator that either mutates the value of a previously perturbed character or mutates the perturbation to a new position. In the first case, the perturbation position remains fixed, and the character is replaced by sampling from a predefined character set. In the second case, we select one unperturbed position to apply a new mutation, while simultaneously restoring one previously perturbed position to its original character.
This two-step policy allows GA to maintain a fixed perturbation budget $k$ while adaptively exploring new locations and values within the mutation space.

\textbf{Heuristic Search.} To perform heuristic search over prompts, we integrate Monte Carlo Tree Search (MCTS) into our attack pipeline. The root nodes of the MCTS tree are initialized using the coarse-grained perturbed prompts generated by GA, which serve as high-potential candidates for further exploration. MCTS then conducts a symbolic search over suffix-level character perturbations, aiming to identify the most semantically disruptive sequence of $m$ special characters that can be appended to each prompt.
The overall procedure follows the standard MCTS paradigm — Selection, Expansion, Evaluation, Simulation, and Backpropagation.

\textbf{Selection.} The root of each tree is initialized with one of the top-$K$ perturbed prompts $x^{\prime}\in\chi_{\mathrm{GA}}$ produced by GA. From the root node, a child is selected at each depth level until a leaf is reached. The path is chosen to minimize the cosine similarity score between the perturbed prompt and the clean prompt $x$. Specifically, each node stores a cumulative similarity-based reward, and selection favors the path with the minimal average similarity over visited children.

\textbf{Expansion.} Given a selected leaf node at depth $d<m$, it is expanded by appending a new character $c_{i}\in\mathcal{C}$ from a predefined character set to the current suffix. This creates a new child node representing a prompt with an extended suffix. The expansion operation at depth $d$ results in a new candidate prompt as Eq.~\eqref{Equ:concat}:
\begin{equation}
x^{\prime\prime}=\mathrm{concat}(x^{\prime},c_1,\ldots,c_d),\quad\mathrm{where~}d\leq m,
\label{Equ:concat}
\end{equation}
\textbf{Evaluation.} In this step, each expanded child node is evaluated with a scalar score that reflects its semantic deviation from the original prompt. 
We still choose cosine similarity as the value and it can be calculated as Eq.~\eqref{Equ:value}:
\begin{equation}
\mathcal{V}(x,x^{\prime\prime})=\cos\left(\tau_\theta(x),\tau_\theta(x^{\prime\prime})\right),
\label{Equ:value}
\end{equation}
The value $V$ serves as the reward to be propagated upward in the tree. 

\textbf{Simulation}. For nodes at depth less than $m$, we simulate full suffixes by randomly sampling additional characters until a complete $m$-character suffix is formed. Multiple such rollouts are executed to explore different completions. Each simulation is independently evaluated via the loss $L$, and the minimum value across all simulations is cached as Eq.~\eqref{Equ:sim}:
\begin{equation}
\mathcal{L}_{sim}^{\min}=\min_{x^{\prime\prime}\in\mathrm{Sim}(x^{\prime})}\mathcal{L}(x,x^{\prime\prime}),
\label{Equ:sim}
\end{equation}
Unlike classical MCTS, which averages path rewards, we explicitly preserve the most effective suffix in simulation to avoid discarding rare but optimal perturbations.

\textbf{Backpropagation}. After evaluation and simulation, we propagate similarity values back up the tree to update parent nodes. Each node $s_i$ in the path $\{s_0,s_1,\ldots,s_l\}$ is updated using the standard visitation count and value accumulation as Eq.~\eqref{Equ:back}:
\begin{equation}
\begin{aligned}
 & N(s_{i})=N(s_i)+1 \\
 & V(s_{i})=\frac{V(s_i)\cdot(N(s_i)-1)+\mathcal{L}}{N(s_i)},
\end{aligned}
\label{Equ:back}
\end{equation}
where $N(s_{i})$ is the number of visits to a node $s$, $V(s_i)$ is the value function (expected return) from the subtree of $s$.
After two stages of attack, we compare three sources of refined prompts: (1) the best perturbation from GA($\mathcal{L}_{\mathrm{GA}}$), (2) the lowest-scoring suffix from the MCTS path($\mathcal{L}_{\mathrm{MCTS}}$) and (3) the best simulation result from rollouts($\mathcal{L}_{sim}^{\min}$). The final perturbed prompt is chosen as Eq.~\eqref{Equ:final}:
\begin{equation}
x^{\text{per}} = \arg\min_{x' \in {\mathcal{X}_{\text{GA}}, \text{MCTS}, \text{Sim}}} \mathcal{L}(x, x'),
\label{Equ:final}
\end{equation}

\section{Experiments}
\begin{figure*}[t]
    \centering
\includegraphics[width=1.0\linewidth]{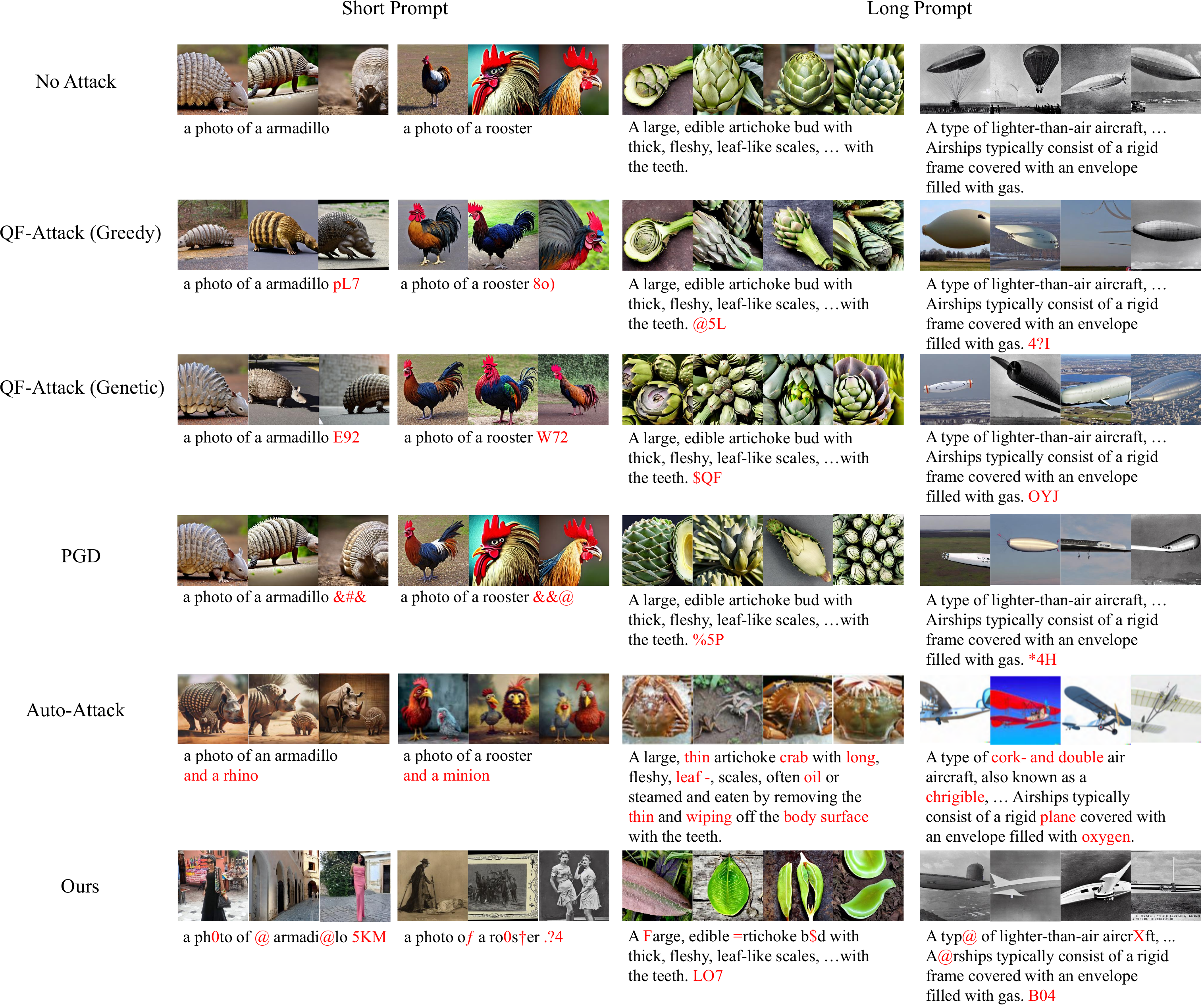}
    \caption{Visual comparison of image outputs under different prompt attack methods on short prompts. Our method generates semantically divergent images.}
    \label{fig:short_attacks}
\end{figure*}
\subsection{Experimental Setup}

\textbf{Implementation Details.}
We adopt the pretrained Stable Diffusion v1.4 as the target generative model for all experiments. Our attack method accesses the pretrained CLIP model (ViT-L/14), which serves as the text encoder for Stable Diffusion. In the stage of selecting root node, the number of perturbed characters $k$ is set to 3, and in the heuristic search phase, the number of perturbed characters $m$ is set to 3. All experiments are conducted on a single NVIDIA RTX 3090 GPU.

\textbf{Baselines.}
We compare CAHS-Attack against both black-box and white-box adversarial attack methods:

QF-Attack (Greedy) and QF-Attack (Genetic): Two query-based black-box attacks introduced in~\cite{zhuang2023pilot}, designed for text-to-image models via iterative clean text prompt modification.

Auto-Attack (ATM) and PGD: White-box baselines that leverage gradient-based optimization to identify adversarial prompts~\cite{du2023stable, zhuang2023pilot}. Despite assuming access to model internals, they serve as upper bounds for comparison.


\textbf{Datasets.}
We evaluate on two prompt datasets introduced in~\cite{du2023stable}:

ImageNet-Short: A template-based set of concise prompts derived from the 1,000 ImageNet-1K classes, each formulated as \textit{"A photo of [CLASS\_NAME]"}.

ImageNet-Long: A more complex prompt set generated via ChatGPT-4, where prompts are syntactically rich and semantically diverse. All prompts are truncated or padded to 77 tokens to match CLIP’s maximum tokenization length.


\textbf{Evaluation Metrics.}
We assess attack performance using three complementary metrics:

Text Similarity (TS): Measures the semantic closeness between clean and adversarial prompts via cosine similarity in the CLIP embedding space. Lower values indicate stronger deviation.

Fréchet Inception Distance (FID)\cite{heusel2017gans}: Quantifies distributional shift between generated images and real images. Higher FID implies greater image degradation or semantic misalignment.

CLIP Score (CS)\cite{hessel2021clipscore}: Computes the alignment between text prompts and generated images using CLIP's vision-language similarity. Lower CLIP scores reflect weaker correspondence, indicating a more successful attack.


Following prior work~\cite{du2023stable}, we generate 50,000 images per evaluation (50 per class) to ensure statistical robustness.

\subsection{Main Results}

We present qualitative and quantitative comparisons between our method and existing baselines on both short and long prompt scenarios.

\textbf{Qualitative Results.}
Fig.\ref{fig:short_attacks} illustrate representative image generation results under six settings: clean (unaltered) prompts, and adversarial prompts generated by CAHS-Attack, QF-Attack (Greedy), QF-Attack (Genetic), Auto-Attack, and PGD. Perturbed characters are highlighted in red for visual clarity. As shown, adversarial prompts crafted by CAHS-Attack lead to significantly altered semantic outputs in the generated images compared to clean prompts, effectively disrupting the intended meaning. 

\begin{table}[ht]
\centering
\caption{Quantitative comparison of different attack methods under different prompt lengths.
We report FID, Text Similarity (TS), and CLIP Score (CS) on short and long prompts. The best results are highlighted in \textbf{bold}}
\label{table1}
\resizebox{1\linewidth}{!}{  
\begin{tabular}{l l c c c}
\hline
\textbf{Prompt} & \textbf{Method} & \textbf{FID~$\uparrow$} & \textbf{TS~$\downarrow$} & \textbf{CS~$\downarrow$} \\
\hline
\multirow{5}{*}{Short} 
& No Attack           & 18.510  & 1.000 & 0.279 \\
& QF-Attack (Greedy)  & 23.889  & 0.721 & 0.276 \\
& QF-Attack (Genetic) & 31.349  & 0.600 & 0.257 \\
& PGD                 & 29.651  & 0.653 & 0.263 \\
& Auto-Attack         & 30.092  & 0.522 & 0.269 \\
& CAHS-Attack (Ours)     & \textbf{118.920} & \textbf{0.185} & \textbf{0.149} \\
\hline
\multirow{5}{*}{Long}
& No Attack           & 17.950  & 1.000 & 0.318 \\
& QF-Attack (Greedy)  & 15.808  & 0.818 & 0.313 \\
& QF-Attack (Genetic) & 16.639  & 0.797 & 0.318 \\
& PGD                 & 15.902  & 0.806 & 0.325 \\
& Auto-Attack         & 16.337  & 0.781 & 0.331 \\
& CAHS-Attack (Ours)     & \textbf{79.325} & \textbf{0.328} & \textbf{0.253} \\
\hline
\end{tabular}
}
\end{table}

\textbf{Quantitative Results.}
Table~\ref{table1} reports the core metrics across all evaluated methods. The results are reported separately for \textit{ImageNet-Short} and \textit{ImageNet-Long} datasets.

Short prompts contain very limited semantic context, making them highly sensitive to even minimal character perturbations. In this setting, the strongest baseline achieves up to 52.2\% TS. In contrast, long prompts provide richer linguistic context and redundancy, requiring more nuanced perturbations to achieve the same degree of semantic drift. In both scenarios, CAHS-Attack outperforms all baselines by a significant margin, achieving 18.5\% TS on short prompts and 32.8\% TS on long prompts, demonstrating both the generality and robustness of our attack framework.

In addition to textual similarity, we also examine the impact of perturbations on the generated image quality and alignment. As the TS decreases (i.e., semantic similarity decreases), we observe a clear degradation in both image fidelity and text-image alignment. Our method achieves a competitive FID of 118.920, a CS of 0.149 on short prompt and FID of 79.325, CS of 0.253 on long prompt, which significantly disrupting semantic correspondence — underscoring its effectiveness as a stealthy and transferable black-box attack.


\begin{table}[ht]
\centering
\tiny
\caption{Ablation study on the necessity of each attack stage. The best results are highlighted in \textbf{bold}}
\label{table2}
\resizebox{1\linewidth}{!}{
\begin{tabular}{l l c c c}
\hline
\textbf{Prompt} & \textbf{Method} & \textbf{FID~$\uparrow$} & \textbf{TS~$\downarrow$} & \textbf{CS~$\downarrow$} \\
\hline
\multirow{3}{*}{Short} 
& GA            & 65.567           & 0.234           & 0.163 \\
& MCTS         & 26.732           & 0.622           & 0.259 \\
& \textbf{CAHS-Attack} & \textbf{118.920} & \textbf{0.185}  & \textbf{0.149} \\
\hline
\multirow{3}{*}{Long} 
& GA            & 29.872           & 0.388           & 0.256 \\
& MCTS          & 15.953           & 0.799           & 0.323 \\
& \textbf{CAHS-Attack} & \textbf{79.325}  & \textbf{0.328}  & \textbf{0.253} \\
\hline
\end{tabular}
}
\end{table}

\subsection{Ablation Study}
\textbf{Effectiveness of Each Module.}
We first evaluate the impact of the root note selector and heuristic search modules when applied independently, compared to the complete pipeline. Results are summarized in Table~\ref{table2}. When using root note selector alone, the attack lacks the local refinement required to maximize embedding deviation. In contrast, heuristic search alone provides limited disruption due to its restricted perturbation space. The full method consistently outperforms both single-stage variants across all metrics, highlighting the complementary nature of global and localized perturbations.

\begin{table}[ht]
\centering
\tiny
\caption{Ablation study on the importance of stage order. The best results are highlighted in \textbf{bold}.}
\label{table3}
\resizebox{1\linewidth}{!}{
\begin{tabular}{l l c c c}
\hline
\textbf{Prompt} & \textbf{Method} & \textbf{FID~$\uparrow$} & \textbf{TS~$\downarrow$} & \textbf{CS~$\downarrow$} \\
\hline
\multirow{2}{*}{Short}
& MCTS-GA & 102.052          & 0.269          & 0.181 \\
& \textbf{GA-MCTS} & \textbf{118.920} & \textbf{0.185} & \textbf{0.149} \\
\hline
\multirow{2}{*}{Long}
& MCTS-GA & 72.314           & 0.355          & 0.285 \\
& \textbf{GA-MCTS} & \textbf{79.325}  & \textbf{0.328} & \textbf{0.253} \\
\hline
\end{tabular}
}
\end{table}

\textbf{Importance of Stage Order.}
To evaluate whether the attack stages are order-sensitive, we reverse the pipeline by applying MCTS globally over the full prompt and GA-based mutation at the suffix. Results are summarized in Table~\ref{table3}. This inverted design significantly underperforms the original configuration. 

\begin{table}[ht]
\centering
\tiny
\caption{Ablation study on the role of adaptive suffix search. The best results are highlighted in \textbf{bold}.}
\label{table4}
\resizebox{1\linewidth}{!}{
\begin{tabular}{l l c c c}
\hline
\textbf{Prompt} & \textbf{Method} & \textbf{FID~$\uparrow$} & \textbf{TS~$\downarrow$} & \textbf{CS~$\downarrow$} \\
\hline
\multirow{2}{*}{Short}
& GA-Suffix & 110.792          & 0.212          & 0.160 \\
& \textbf{CAHS-Attack} & \textbf{118.920} & \textbf{0.185} & \textbf{0.149} \\
\hline
\multirow{2}{*}{Long}
& GA-Suffix & 74.362           & 0.367          & 0.257 \\
& \textbf{CAHS-Attack} & \textbf{79.325}  & \textbf{0.328} & \textbf{0.253} \\
\hline
\end{tabular}
}
\end{table}

\textbf{Role of Adaptive Suffix Search.}
We further assess the necessity of adaptive suffix construction by comparing heuristic search with a variant that appends a fixed 3-character token (e.g., “\#\#\#”) to each perturbed prompt. Results are summarized in Table~\ref{table4}. While both settings maintain consistent prompt length, the fixed-suffix variant yields significantly lower attack success rates. This demonstrates that the effectiveness of heuristic search arises not from length variation but from its ability to discover semantically disruptive suffixes tailored to the upstream prompt structure.

\section{Conclusion}
In this paper, we proposed CAHS-Attack, a black-box adversarial attack framework that leverages CLIP-aware heuristic search to uncover vulnerabilities in Stable Diffusion models. CAHS-Attack introduce a genetic algorithm–based root node selector module that identifies high-potential adversarial prompts to seed the MCTS process. Furthermore, we enhance the effectiveness of MCTS by retaining the most semantically disruptive outcome from each simulation rollout, ensuring locally optimal adversarial prompt are preserved. Extensive experiments on both short and long text prompts across diverse semantic categories demonstrate that CAHS-Attack achieves state-of-the-art performance under black-box conditions. Our studies also reveal that the core vulnerability of Stable Diffusion stems from the inherent fragility of its CLIP-based text encoder.

\bibliographystyle{IEEEtran}
\bibliography{IEEEfull}

@inproceedings{rombach2022high,
  title={High-resolution image synthesis with latent diffusion models},
  author={Rombach, Robin and Blattmann, Andreas and Lorenz, Dominik and Esser, Patrick and Ommer, Bj{\"o}rn},
  booktitle={Proceedings of the IEEE/CVF Conference on Computer Vision and Pattern Recognition},
  pages={10684--10695},
  year={2022}
}

@inproceedings{xie2025progressive,
  title={Progressive autoregressive video diffusion models},
  author={Xie, Desai and Xu, Zhan and Hong, Yicong and Tan, Hao and Liu, Difan and Liu, Feng and Kaufman, Arie and Zhou, Yang},
  booktitle={Proceedings of the Computer Vision and Pattern Recognition Conference},
  pages={6322--6332},
  year={2025}
}

@misc{stablility2023stable,
  author="Stability-AI",
  title="Stable Diffusion Public Release",
  year="2023",
  howpublished="\url{https://stability.ai/blog/stable-diffusion-public-release}",
  note="Accessed on 2023-05-17",
}

@misc{flux2024flux,
  author="flux",
  title="FLUX",
  year="2024",
  howpublished="\url{https://flux1.ai/}",
  note="Accessed on 2024-08-01",
}

@inproceedings{voleti2024sv3d,
  title={Sv3d: Novel multi-view synthesis and 3d generation from a single image using latent video diffusion},
  author={Voleti, Vikram and Yao, Chun-Han and Boss, Mark and Letts, Adam and Pankratz, David and Tochilkin, Dmitry and Laforte, Christian and Rombach, Robin and Jampani, Varun},
  booktitle={European Conference on Computer Vision},
  pages={439--457},
  year={2024},
  organization={Springer}
}

@article{heusel2017gans,
  title={Gans trained by a two time-scale update rule converge to a local nash equilibrium},
  author={Heusel, Martin and Ramsauer, Hubert and Unterthiner, Thomas and Nessler, Bernhard and Hochreiter, Sepp},
  journal={Advances in neural information processing systems},
  volume={30},
  year={2017}
}

@inproceedings{hessel2021clipscore,
  title={Clipscore: A reference-free evaluation metric for image captioning},
  author={Hessel, Jack and Holtzman, Ari and Forbes, Maxwell and Bras, Ronan Le and Choi, Yejin},
  booktitle={Proceedings of the 2021 Conference on Empirical Methods in Natural Language Processing},
  year={2021}
}

@article{yang2025mmada,
  title={Mmada: Multimodal large diffusion language models},
  author={Yang, Ling and Tian, Ye and Li, Bowen and Zhang, Xinchen and Shen, Ke and Tong, Yunhai and Wang, Mengdi},
  journal={arXiv preprint arXiv:2505.15809},
  year={2025}
}

@inproceedings{zhuang2023pilot,
  title={A pilot study of query-free adversarial attack against stable diffusion},
  author={Zhuang, Haomin and Zhang, Yihua and Liu, Sijia},
  booktitle={Proceedings of the IEEE/CVF Conference on Computer Vision and Pattern Recognition},
  pages={2385--2392},
  year={2023}
}

@article{du2023stable,
  title={Stable diffusion is unstable},
  author={Du, Chengbin and Li, Yanxi and Qiu, Zhongwei and Xu, Chang},
  journal={Advances in Neural Information Processing Systems},
  volume={36},
  pages={58648--58669},
  year={2023}
}

@inproceedings{yang2024mma,
  title={Mma-diffusion: Multimodal attack on diffusion models},
  author={Yang, Yijun and Gao, Ruiyuan and Wang, Xiaosen and Ho, Tsung-Yi and Xu, Nan and Xu, Qiang},
  booktitle={Proceedings of the IEEE/CVF Conference on Computer Vision and Pattern Recognition},
  pages={7737--7746},
  year={2024}
}

@inproceedings{huang2025perception,
  title={Perception-guided jailbreak against text-to-image models},
  author={Huang, Yihao and Liang, Le and Li, Tianlin and Jia, Xiaojun and Wang, Run and Miao, Weikai and Pu, Geguang and Liu, Yang},
  booktitle={Proceedings of the AAAI Conference on Artificial Intelligence},
  volume={39},
  number={25},
  pages={26238--26247},
  year={2025}
}

@article{ho2020denoising,
  title={Denoising diffusion probabilistic models},
  author={Ho, Jonathan and Jain, Ajay and Abbeel, Pieter},
  journal={Advances in neural information processing systems},
  volume={33},
  pages={6840--6851},
  year={2020}
}

@inproceedings{wang2024instancediffusion,
  title={Instancediffusion: Instance-level control for image generation},
  author={Wang, Xudong and Darrell, Trevor and Rambhatla, Sai Saketh and Girdhar, Rohit and Misra, Ishan},
  booktitle={Proceedings of the IEEE/CVF conference on computer vision and pattern recognition},
  pages={6232--6242},
  year={2024}
}

@article{wen2023hard,
  title={Hard prompts made easy: Gradient-based discrete optimization for prompt tuning and discovery},
  author={Wen, Yuxin and Jain, Neel and Kirchenbauer, John and Goldblum, Micah and Geiping, Jonas and Goldstein, Tom},
  journal={Advances in Neural Information Processing Systems},
  volume={36},
  pages={51008--51025},
  year={2023}
}

@inproceedings{corneanu2024latentpaint,
  title={Latentpaint: Image inpainting in latent space with diffusion models},
  author={Corneanu, Ciprian and Gadde, Raghudeep and Martinez, Aleix M},
  booktitle={Proceedings of the IEEE/CVF winter conference on applications of computer vision},
  pages={4334--4343},
  year={2024}
}

@article{chen2025multidiffeditattack,
  title={MultiDiffEditAttack: A Multi-Modal Black-Box Jailbreak Attack on Image Editing Models},
  author={Chen, Peihong and Chen, Feng and Guo, Lei},
  journal={Electronics},
  volume={14},
  number={5},
  pages={899},
  year={2025},
  publisher={MDPI AG}
}

@article{zhao2025inception,
  title={Inception: Jailbreak the memory mechanism of text-to-image generation systems},
  author={Zhao, Shiqian and Liu, Jiayang and Li, Yiming and Hu, Runyi and Jia, Xiaojun and Fan, Wenshu and Li, Xinfeng and Zhang, Jie and Dong, Wei and Zhang, Tianwei and others},
  journal={arXiv preprint arXiv:2504.20376},
  year={2025}
}

@article{ma2024jailbreaking,
  title={Jailbreaking prompt attack: A controllable adversarial attack against diffusion models},
  author={Ma, Jiachen and Li, Yijiang and Xiao, Zhiqing and Cao, Anda and Zhang, Jie and Ye, Chao and Zhao, Junbo},
  journal={arXiv preprint arXiv:2404.02928},
  year={2024}
}

@inproceedings{liu2024latent,
  title={Latent guard: a safety framework for text-to-image generation},
  author={Liu, Runtao and Khakzar, Ashkan and Gu, Jindong and Chen, Qifeng and Torr, Philip and Pizzati, Fabio},
  booktitle={European Conference on Computer Vision},
  pages={93--109},
  year={2024},
  organization={Springer}
}

@article{yang2024guardt2i,
  title={Guardt2i: Defending text-to-image models from adversarial prompts},
  author={Yang, Yijun and Gao, Ruiyuan and Yang, Xiao and Zhong, Jianyuan and Xu, Qiang},
  journal={Advances in neural information processing systems},
  volume={37},
  pages={76380--76403},
  year={2024}
}

@inproceedings{yang2024sneakyprompt,
  title={Sneakyprompt: Jailbreaking text-to-image generative models},
  author={Yang, Yuchen and Hui, Bo and Yuan, Haolin and Gong, Neil and Cao, Yinzhi},
  booktitle={2024 IEEE symposium on security and privacy (SP)},
  pages={897--912},
  year={2024},
  organization={IEEE}
}

@article{xiong2025prompt,
  title={Prompt suffix-attack against text-to-image diffusion models},
  author={Xiong, Siyun and Du, Yanhui and Wang, Zhuohao and Sun, Peiqi},
  journal={Neurocomputing},
  volume={630},
  pages={129659},
  year={2025},
  publisher={Elsevier}
}

@article{jia2024improved,
  title={Improved techniques for optimization-based jailbreaking on large language models},
  author={Jia, Xiaojun and Pang, Tianyu and Du, Chao and Huang, Yihao and Gu, Jindong and Liu, Yang and Cao, Xiaochun and Lin, Min},
  journal={arXiv preprint arXiv:2405.21018},
  year={2024}
}

@inproceedings{huang2024personalization,
  title={Personalization as a shortcut for few-shot backdoor attack against text-to-image diffusion models},
  author={Huang, Yihao and Juefei-Xu, Felix and Guo, Qing and Zhang, Jie and Wu, Yutong and Hu, Ming and Li, Tianlin and Pu, Geguang and Liu, Yang},
  booktitle={Proceedings of the AAAI Conference on Artificial Intelligence},
  year={2024}
}

@article{liu2025token,
  title={Token-level constraint boundary search for jailbreaking text-to-image models},
  author={Liu, Jiangtao and Wang, Zhaoxin and Wang, Handing and Tian, Cong and Jin, Yaochu},
  journal={arXiv preprint arXiv:2504.11106},
  year={2025}
}

@article{gao2024hts,
  title={HTS-Attack: Heuristic Token Search for Jailbreaking Text-to-Image Models},
  author={Gao, Sensen and Jia, Xiaojun and Huang, Yihao and Duan, Ranjie and Gu, Jindong and Bai, Yang and Liu, Yang and Guo, Qing},
  journal={arXiv preprint arXiv:2408.13896},
  year={2024}
}

@article{guo2024cold,
  title={Cold-attack: Jailbreaking llms with stealthiness and controllability},
  author={Guo, Xingang and Yu, Fangxu and Zhang, Huan and Qin, Lianhui and Hu, Bin},
  journal={arXiv preprint arXiv:2402.08679},
  year={2024}
}

@article{le2022perturbations,
  title={Perturbations in the wild: Leveraging human-written text perturbations for realistic adversarial attack and defense},
  author={Le, Thai and Lee, Jooyoung and Yen, Kevin and Hu, Yifan and Lee, Dongwon},
  journal={arXiv preprint arXiv:2203.10346},
  year={2022}
}

\end{document}